\documentclass{article}
\usepackage{graphicx} 
\usepackage{blindtext}
\usepackage{amsmath}
\usepackage{amsfonts}
\usepackage{amssymb}
\usepackage{amsthm}
\usepackage{braket}
\usepackage{graphicx}
\usepackage{tikz-cd}
\usepackage{caption}
\usepackage{float}
\usepackage[colorlinks=true, allcolors=blue]{hyperref}
\usepackage[dvipsnames]{xcolor}
\usepackage{mathrsfs}
\usepackage{authblk}

\usepackage[numbers,sort&compress]{natbib}

\newcommand{\Z}{{\mathbb{Z}}}
\newcommand{\R}{{\mathbb{R}}}

\newcommand{\beq}{\begin{equation}}
\newcommand{\eeq}{\end{equation}}
\newcommand{\beqn}{\begin{eqnarray}}
\newcommand{\eeqn}{\end{eqnarray}}

\newcommand{\tphi}{{\tilde{\phi}}}
\newcommand{\tchi}{{\tilde{\chi}}}
\newcommand{\tn}{{\tilde{n}}}

\newcommand{\lr}[1]{\lfloor #1 \rceil}
\newcommand{\TP}{{T_{\text{P}}}}
\newcommand{\TAP}{{T_{\text{AP}}}}

\title{1+1d Lattice Dirac Fermions from Non-Onsite Vector and Axial Symmetries}

\author[1]{Tabin Dharanikota}
\author[1]{Lukasz Fidkowski}
\affil[1]{Department of Physics, \protect\\ University of Washington, Seattle, WA 98195, USA}
\date{August 2026}

\begin{document}

\maketitle

\begin{abstract}
We construct an exactly solvable Hamiltonian lattice model realizing a 1+1d Dirac fermion, with exact microscopic vector and axial vector $U(1)$ symmetries.  The mixed anomaly between these is accommodated by the not-on-site action of the symmetries.  Our Hilbert space is a ${\mathbb{Z}}_2$-graded tensor product of local ${\mathbb{Z}}_2$-graded Hilbert spaces which include infinite dimensional rotor degrees of freedom.  The Hamiltonian becomes manifestly exactly solvable after a locality-preserving unitary mapping to an equivalent fermionic Villain Hilbert space.  Our construction also allows an exactly solvable realization of interacting fermionic Luttinger liquids.  At the free Dirac fixed point, our Hamiltonian contains irrelevant interactions, which we compute to leading order.

\end{abstract}

\section{Introduction}

Bosonization in $1+1$d is an exact duality between a $1+1$d Dirac fermion and a $c=1$ compact boson at a special value of the compactification radius.  An important role in this duality is played by the $U(1)_V$ and $U(1)_A$ symmetries, which on the fermionic side are interpreted as the usual vector and axial vector symmetries of the Dirac fermion, and on the bosonic side correspond to compact momentum and winding respectively.  It is well known that $U(1)_V$ and $U(1)_A$ have a mixed anomaly, precluding any lattice realization of either the bosonic or fermionic theory with both symmetries realized onsite. For an overview on bosonization, see \cite{delft_bosonization_1998, stone1994bosonization, senechal_introduction_2004}. 

Recently, however, it has been shown that the compact boson can be realized on the spatial lattice in an exactly solvable modified Villain Hamiltonian with both $U(1)_V$ and $U(1)_A$ as exact symmetries, at the expense of making $U(1)_A$ not-on-site, and introducing infinite dimensional `rotor' degrees of freedom \cite{cheng_lieb-schultz-mattis_2023, seifnashri_exactly_2026, thorngren_chiral_2026} (see also \cite{fidkowski_non-invertible_2025, lu_lattice_2026} for analogous work in $3+1$d, and \cite{villain_theory_1975, sulejmanpasic_abelian_2019, gorantla_modified_2021, baig_bosonization_2026, fazza_lattice_2023} for recent advances related to the modified Villain model; in particular ref. \cite{baig_bosonization_2026} concerns a spacetime lattice formulation of chiral fermion operators).  This makes it natural to wonder whether a fermionic lattice Hamiltonian can be constructed which at low energies flows to a $1+1$d Dirac fermion, with exact microscopic, albeit not-on-site $U(1)_V$ and $U(1)_A$ symmetries.  The not-on-siteness of these symmetries, together with the interactions allow one to evade the Nielsen-Ninomiya theorem \cite{nielsen_no-go_1981}, which is fundamentally a free fermion / quadratic Hamiltonian argument (see also \cite{Chatterjee_2025} for a different approach that evades this theorem by enlarging the symmetry algebra in the UV).  More generally, one can also wonder whether the field-theoretic bosonization duality can be extended to a lattice based duality that maps these two Hamiltonians onto each other, and intertwines the $U(1)_V$ and $U(1)_A$ symmetries.

In this work, we construct an exactly solvable lattice Hamiltonian, with exact $U(1)_V$ and $U(1)_A$ symmetries, that realizes a Dirac fermion at low energies.  In fact, our construction allows us to realize the entire Luttinger liquid universality class of interacting fermionic systems in an exactly solved model.  Our Hamiltonian acts on a fermionic tensor product Hilbert space consisting of Majorana degrees of freedom $\gamma_r, \gamma'_r$ for each lattice site $r$, together with bosonic degrees of freedom ${\tilde{\phi}}_r \in \R / \Z$ and half-integer valued degrees of freedom ${\tilde{n}}_{r,r+1}$ on the edges, which can also be viewed as rotors in dual variables.  In this Hilbert space we impose the constraint $i\gamma_r \gamma'_r \exp\left(-\frac{1}{2} \frac{d}{d{\tilde{\phi_r}}}\right) = 1$, which may be viewed as forcing the odd angular momentum states of the $\phi$ rotor to be fermionic.  Despite the inclusion of these bosonic rotor degrees of freedom and the constraint, the entire Hilbert space is still a $\Z_2$-graded tensor product of local $\Z_2$-graded Hilbert spaces with both bosonic and fermionic sectors - i.e. a fermionic system.

There exists a useful alternative presentation of this Hilbert space in terms of the same Majorana fermions, real-valued variables $\phi_r$, and half-integer valued degrees of freedom $n_{r,r+1}$ with a fermionic Villain condition imposed.  The fermionic Villain condition is that all wavefunctions must be invariant under the operator
\begin{align}
\exp\left(-\frac{1}{2}\frac{d}{d\phi_r}\right) \exp\left(i \chi_{r-1,r} - i \chi_{r,r+1}\right) i\gamma_r\gamma'_r
\end{align}
where (suppressing subscripts for clarity) $\exp\left(i \chi\right)$ is the dual variable to $n$: $\exp\left(i \chi \right)|n\rangle = |n+\frac{1}{2}\rangle$.  These two presentations of the Hilbert space are related by a locality-preserving algebra automorphism, which we call the `Villain disentangler' \cite{thorngren_chiral_2026}, because it maps the fermionic Villain condition above to the onsite constraint $i\gamma_r \gamma'_r \exp\left(-\frac{1}{2} \frac{d}{d{\tilde{\phi_r}}}\right) = 1$.  While one can in principle work in either the tensor product or the Villain Hilbert space, we find it more convenient to work in the Villain one, and then translate the results to the tensor product one.

We construct a Hamiltonian on this fermionic Villain Hilbert space, closely related to that of ref. \cite{cheng_lieb-schultz-mattis_2023}, which we claim realizes a single Dirac fermion at low energies, with exact $\left(U(1)_V \times U(1)_A\right) / \Z_2$ symmetry.  The exact solvability of our Hamiltonian is closely related to that of ref. \cite{cheng_lieb-schultz-mattis_2023} - the Villain field $n$ effectively has no dynamics, reducing the problem to harmonic oscillators, modulo zero modes and winding modes which have to be treated carefully.  In fact, our Hamiltonian can be exactly mapped to that of ref. \cite{cheng_lieb-schultz-mattis_2023} via an exact Jordan-Wigner-like transformation defined in our lattice Hilbert space (see also ref. \cite{baig_bosonization_2026} for an approach involving bosonization on the spacetime lattice).  At low energies, this transformation reduces to the usual field theoretic bosonization duality.  In particular, the solitons carrying the units of charge and winding that correspond to chiral fermions are indeed fermion parity odd states in our Hilbert space.  By tuning the ratio of the kinetic and $XY$ coupling for $\phi_r$ in our Hamiltonian, we can also tune away from the free fermion fixed point, and realize an interacting Luttinger liquid.  Our Hilbert space also has the virtue of allowing both periodic and anti-periodic conditions on the fermions to be imposed naturally, with the resulting discrete translation operator $T$, under which the Hamiltonian is invariant, satisfying $T^N = (-1)^F$ in the anti-periodic case.

A natural question in any such lattice construction is: what is the local odd operator that creates a chiral fermion?  Indeed, the harmonic oscillator nature of our Hamiltonian and its close connection to bosonization suggest that while writing down chiral particle-hole operators should be easy - these correspond to harmonic excitations of the oscillator modes - writing down a fermionic operator is more involved, as it must have non-trivial charge under $U(1)_V$ and $U(1)_A$.  To proceed, we first construct a non-local operator that inserts compact momentum ($U(1)_V$ charge) and winding ($U(1)_A$ charge), with periodic boundary conditions, in the natural way suggested by the harmonic oscillator problem.  We show that this operator is indeed odd under fermion parity, and takes eigenstates of our Hamiltonian to other eigenstates.  However, it should not be thought of as the creation operator for a Fock space mode: indeed, its action is not simply to change the occupation number of a particular fermionic mode.  Rather, it is an operator which uniformly moves all of the modes on a chiral - say right-moving - branch of the dispersion up by one, effectively pulling in a fermion from the Dirac sea.  The fact that this is a unitary operator which can pull in unlimited charge is allowed in our setting by our use of infinite dimensional local rotor Hilbert spaces.

One may attempt to localize this operator to produce soliton localized in a finite interval of length $\ell$.  One can in fact do this for any $\ell$, down to $\ell = 2a$, where $a$ is the lattice spacing, but the result, despite having the correct, well defined, $U(1)_V$ and $U(1)_A$ charges of a chiral fermion, is not a chiral operator.  Instead, one may think of the excitation it creates as a chiral fermion together with particle-hole excitations of both chiralities.  For large $\ell / a$, at the free Dirac fermion fixed point, one can make this operator approximately chiral, with the chirality violation being exponentially small in $\ell / a$. 
When we tune to an interacting Luttinger liquid, any way of constructing a localized operator that inserts these quantum numbers of $U(1)_V$ and $U(1)_A$ leads to a non-chiral excitation, one that spreads in both directions, consistent with the conformal field theory (CFT) expectation.
Due to the interacting nature of our Hamiltonian, we expect it to contain irrelevant corrections, even at the free fermion Dirac fixed point.  We show that these corrections first arise with scaling dimension $4$, and are due to deviations away from linearity in the dispersion of the bosonic mode describing the particle-hole excitations.  Using the exact solution of our Hamiltonian we compute the coefficient of the leading irrelevant correction.

We believe this work will elucidate the broader problem of how to put chiral fermions on the lattice. As dictated by the Nielsen-Ninomiya theorem \cite{nielsen_no-go_1981}, naive discretization of continuum chiral theories leads to the ``fermion doubling" problem.  There exist several ways to lift the assumptions of the theorem while still retaining a readily analyzable model.  One strategy is known as the ``domain wall fermion" approach, where chiral fermions appear as boundary modes of a symmetry protected topological phase \cite{kaplan_method_1992,kaplan_chiral_2012}.  The cost of this approach is the requirement of analyzing a system outside of the original spacetime dimension. 
When symmetry anomalies all cancel, an alternate but related approach becomes possible, which couples a finite slab domain wall fermion realization with a strong interaction that symmetrically gaps out the fermions on one side of the slab \cite{eichten_chiral_1986, wang_solution_2019, wang_non-perturbative_2023, wang_symmetric_2022}.  However, this approach is often complicated by the requirement of strong interactions that are outside of the regime of validity of the continuum field theory.

Through our fermionization of an exactly solvable model whose symmetries capture the anomaly data in the UV which (through the use of disentanglers) can readily be made microscopic and hence gauged using standard lattice gauge theory techniques, we arrive at a theory that avoids many drawbacks of previous attempts regulate chiral fermionic systems.

During the completion of this work, we learned about related work by Zhiyao Lu and Shu-Heng Shao, `Fermionic Villain model with exact lattice chiral symmetries', and coordinated the arXiv submission with them.

\section{Fermionic Villain and tensor product Hilbert spaces}

We will now construct our fermionic Hilbert space.  We will work on a ring with $N$ lattice sites, labeled $0,\ldots, N-1$.  We will always take site indices modulo $N$.  We will present two different constructions of a $\Z_2$-graded Hilbert space, and then specify the equivalence between them via a so-called `Villain disentangler' $C$.

\begin{figure}[htbp]
    \centering
    \includegraphics[width=0.6\textwidth]{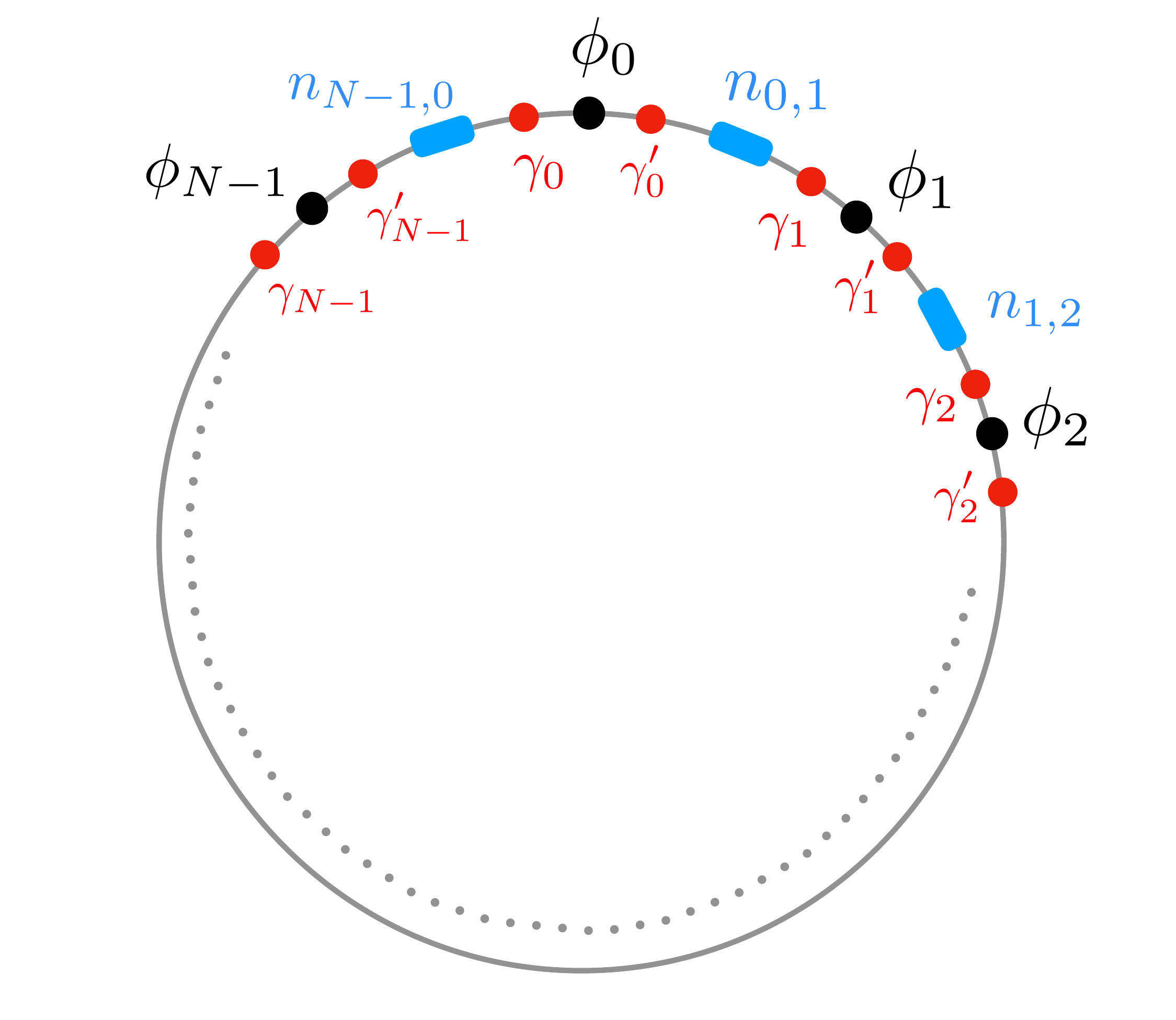}
    \caption{Degrees of freedom of the fermionic Villain model.  The bosonic degrees of freedom are $\phi_r \in \R$ and $n_{r,r+1} \in \Z/2$.  The fermionic degrees of freedom are Majorana fermions $\gamma_r, \gamma'_r$.  There is a fermionic Villain condition (eq. \ref{eq:Villain_condition}) imposed on all of the states in the Hilbert space.  In the (equivalent) graded tensor product Hilbert space, $\phi_r$ are replaced with $\tphi_r \in \R / \Z$, and $n_{r,r+1}$ by $\tn_{r,r+1} \in \Z/2$, and the Villain condition maps onto the constraint that odd angular momentum states of the $\tphi_r$ rotor are fermion parity odd ($i\gamma_r \gamma_r' = -1$).  }
    \label{fig:my-figure}
\end{figure}

{\bf{Fermionic Villain:}}
Let $r =0,\ldots, N-1$ label the lattice sites.  The degrees of freedom are then $\phi_r \in \R$, $n_{r,r+1} \in \Z/2$, together with Majorana fermions $\gamma_r, \gamma'_r$.  Letting $\exp(i\chi_{r,r+1})$ be the dual field to $n_{r,r+1}$, i.e. one satisfying $\exp(i\chi_{r,r+1})|n\rangle = |n+\frac{1}{2}\rangle$, we also impose a fermionic Villain condition:
\begin{align} \label{eq:Villain_condition}
\exp\left(-\frac{1}{2}\frac{d}{d\phi_r}\right) \exp\left(i \chi_{r-1,r} - i \chi_{r,r+1}\right) i\gamma_r\gamma'_r
\end{align}
One may alternatively view this Hilbert space as arising from a two-step construction.  First, we build the bosonic Hilbert space based on $\phi_r$ and $n_{r,r+1}$, with the square of the condition in eq. \ref{eq:Villain_condition} imposed: this is just the ordinary Villain constraint $\exp\left(-\frac{d}{d\phi_r}\right) \exp\left(2i \chi_{r-1,r} - 2i \chi_{r,r+1}\right)=1$, which amounts to the requirement that wavefunctions $\Psi(\phi,n)$ satisfy
\begin{align*}
\Psi(\phi+s,n+ds) = \Psi(\phi,n)
\end{align*}
for any $s \in C^1(S^1,\Z)$ (see ref. \cite{thorngren_chiral_2026} for a review of the chochain notation).  Second, we tensor this bosonic Hilbert space with the $\Z_2$-graded Hilbert space of the Majorana fermions $\gamma_r, \gamma'_r$, and impose eq. \ref{eq:Villain_condition}.  The advantage of this formulation is that it allows one to easily construct a normalized inner product on this Hilbert space.

We will define the fermion parity
\begin{align*}
(-1)^F = \prod_{r=0}^{N-1} \left(i\gamma_r \gamma'_r\right).
\end{align*}
It will be useful to enumerate a generating set for the local observables on this Hilbert space.  First, let us consider fermion parity even observables.  We claim that these can be generated by
\begin{itemize} \label{eq:generator_list_Villain}
\item[\tiny $\bullet$] $\exp(4\pi i \phi_r)$;
\item[\tiny $\bullet$] $\phi_{r+1}-\phi_r - n_{r,r+1}$;
\item[\tiny $\bullet$] $\frac{d}{d\phi_r}$;
\item[\tiny $\bullet$] $\exp(i\chi_{r,r+1})$;
\item[\tiny $\bullet$] $i\gamma'_r \gamma_{r+1}\exp(2\pi i (\phi_{r+1}-\phi_r))$.
\end{itemize}

Finally, the addition of a single odd fermion parity operator, say $\exp(\pi i \phi_r) \gamma_r$, generates the whole algebra.  The fact that these generate the whole algebra will be easiest to see after we establish the equivalence to the graded tensor product Hilbert space.

{\bf{Fermionic graded tensor product:}}
Again, let $r = 0,\ldots,N-1$ denote a lattice site.  Our degrees of freedom are now bosonic rotors $\tphi_r \in \R / \Z$, discrete $\tn_{r,r+1} \in \Z / 2$ (which may be viewed as dual rotors), and Majorana fermions $\gamma_r, \gamma'_r$.  On these we impose the constraint 
\begin{align} \label{eq:tensor_prod_constraint}
\exp\left(-\frac{1}{2} \frac{d}{d\tilde\phi_r}\right)i\gamma_r \gamma'_r = 1
\end{align}
which amounts to the requirement that odd angular momentum states of the $\phi_r$ rotors be fermion parity odd.  Thus, the Hilbert space here is explicitly a graded tensor product of local $\Z_2$-graded Hilbert spaces.  This fact allows us to immediately write down a set of local generators for the fermion parity even sub-algebra:
\begin{itemize} \label{eq:generator_list_TP}
\item[\tiny $\bullet$] $\exp(4\pi i \tphi_r)$;
\item[\tiny $\bullet$] $\tn_{r,r+1}$;
\item[\tiny $\bullet$] $\frac{d}{d\tphi_r}$;
\item[\tiny $\bullet$] $\exp(i\tchi_{r,r+1})$;
\item[\tiny $\bullet$] $i\gamma'_r \gamma_{r+1}\exp(2\pi i (\tphi_{r+1}-\tphi_r))$.
\end{itemize}
Indeed, $\tn_{r,r+1}$ and $\exp(i\tchi_{r,r+1})$ generate the algebra of the bosonic dual rotors, and $\exp(4\pi i \tphi_r)$ and $\frac{d}{d\tphi_r}$ generate the even algebra of the $\phi_r$ rotors.  Since $\gamma_r \exp(2\pi i \tphi_r)$ generates the odd algebra of the $\phi_r$ rotor (recall that $i\gamma_r \gamma'_r = \exp\left(\frac{1}{2}\frac{d}{d\tphi_r}\right)$ is already in the even algebra), the last set of terms above, namely $i\gamma'_r \gamma_{r+1}\exp(2\pi i (\tphi_{r+1}-\tphi_r))$, can be thought of as nearest neighbor hopping operators of fermion parity.  Now, any overall even operator can be built as a linear combination of local graded tensor products, with an even number of terms in each graded tensor factor being fermion parity odd. 
Each such graded tensor factor can be expressed as some product of the hopping operators, times an ordinary tensor product of local even operators, which are already generated by the previous four terms on the list.  Finally, the addition of any single odd operator, such as $i\gamma_r \exp(2\pi i \tphi_r)$, generates the whole algebra.

{\bf{Fermionic Villain disentangler:}}
As in the bosonic case in ref. \cite{thorngren_chiral_2026}, we construct a `disentangler' $C$, which is a locality preserving isomorphism from the fermionic Villain to the fermionic graded tensor product algebra.  One may informally think of this as a unitary from the Villain Hilbert space to the graded tensor product one which acts only on the $\phi_r, n_{r,r+1}$ degrees of freedom (i.e. acts as the identity on the Fock space of the $\gamma_r,\gamma'_r$) by
\begin{align*}
    C|\phi,n\rangle = |\phi,n-\frac{1}{2}\lr{2d\phi}\rangle. \tag{informal}
\end{align*}
Here we have used the cochain notation, and the `nearest integer' function $\lr{x}$, which is just defined as the integer closest to the real number $x$.  In fact, this is just the Villain disentangler of \cite{thorngren_chiral_2026}, with $\phi, n$ rescaled by a factor of $2$.  As shown there, this maps the $\exp\left(-\frac{1}{2}\frac{d}{d\phi_r}\right) \exp\left(i \chi_{r-1,r} - i \chi_{r,r+1}\right)$ to $\exp\left(-\frac{1}{2}\frac{d}{d\phi_r}\right)$, and hence maps the Villain condition (eq. \ref{eq:Villain_condition}) to the constraint in the tensor product Hilbert space (eq. \ref{eq:tensor_prod_constraint}).

More formally, we define the fermionic Villain disentangler as an isomorphism of local operator algebras (here we abuse notation slightly and use $C$ to denote both the unitary, and the associated operator algebra map induced by conjugation):

\[
\begin{array}{rcl}
C:\qquad
\exp(4\pi i \phi_r) &\longmapsto& \exp(4\pi i \tphi_r),\\
\phi_{r+1}-\phi_r - n_{r,r+1} &\longmapsto& \tphi_{r+1}-\tphi_r - \frac{1}{2}\lr{2\tphi_{r+1}-2\tphi_r}-\tn_{r,r+1},\\
\frac{d}{d\phi_r} &\longmapsto& \frac{d}{d\tphi_r},\\
\exp(i\chi_{r,r+1}) &\longmapsto& \exp(i\tchi_{r,r+1}),\\
i\gamma'_r \gamma_{r+1}\exp(2\pi i (\phi_{r+1}-\phi_r)) &\longmapsto& i\gamma'_r \gamma_{r+1}\exp(2\pi i (\tphi_{r+1}-\tphi_r)).
\end{array}
\]
We see that all of the local generating operators in the fermionic Villain Hilbert space map to the corresponding ones in the fermionic tensor product one, except $d\phi-n$, which has a slightly more complicated image.  We also note that $\tphi_{r+1}-\tphi_r - \frac{1}{2}\lr{2\tphi_{r+1}-2\tphi_r}$ is invariant under individual shifts of any $\phi_r$ by any half-integer, and hence can be expressed as a series in $\exp(4\pi i m \phi_r), \exp(4\pi i m' \phi_{r+1})$, which are powers of the generators listed above.  This defines the isomorphism $C$.

One consequence of the above definition is that
\[
\begin{array}{rcl}
C^{-1}:\qquad
\tn_{r,r+1} &\longmapsto& n_{r,r+1}-\frac{1}{2}\lr{2\phi_{r+1}-2\phi_r}
\end{array}
\]
To explicitly express the operator on the right hand side of the above equation in terms of the local generators we wrote down earlier, we note that
\begin{align*}
n&_{r,r+1} -\frac{1}{2}\lr{2\phi_{r+1}-2\phi_r} \\ &= \left(\phi_{r+1}-\phi_r - \frac{1}{2}\lr{2\phi_{r+1}-2\phi_r}\right) - \left(\phi_{r+1}-\phi_r - n_{r,r+1}\right)
\end{align*}
and again use the fact that the first bracketed expression on the right hand side is invariant under half-integer shifts of the $\phi_r$.

\section{Symmetries}
{\bf{Translation symmetry:}}
Let us first define the lattice translation symmetry operator.  There are actually two different translation operators we will define, $\TP$ and $\TAP$, depending on if the boundary conditions on the fermions are periodic or anti-periodic, respectively.  Both of them will act by
incrementing the lattice index by $1$ (modulo $N$) on all the bosonic generators listed above, e.g.
\begin{align*}
\TP \exp(4\pi i \phi_r) \TP^{-1} &= \exp(4\pi i \phi_{r+1}) \\
\TAP \exp(4\pi i \phi_r) \TAP^{-1} &= \exp(4\pi i \phi_{r+1})
\end{align*}
On the fermions $\gamma_r, \gamma'_r$, the translation operators will act as follows:
\[
\TP \gamma_r \TP^{-1} = \gamma_{r+1} \text{ for } r=0,\ldots,N-1
\]
\[
\TP \gamma'_r \TP^{-1} = \gamma'_{r+1} \text{ for } r=0,\ldots,N-1
\]
and
\[
\TAP \gamma_r \TAP^{-1} =
\begin{cases}
\gamma_{r+1}, & \text{if } r=0,\ldots,N-2,\\
-\gamma_{r+1}, & \text{if } r=N-1.
\end{cases}
\]
\[
\TAP \gamma'_r \TAP^{-1} =
\begin{cases}
\gamma'_{r+1}, & \text{if } r=0,\ldots,N-2,\\
-\gamma'_{r+1}, & \text{if } r=N-1.
\end{cases}
\]
Of course these operators are not individually invariant under the Villain constraints, but the above definition uniquely defines the action on the Villain-invariant operators.  Note that $\TP^N = 1$, and $\TAP^N = (-1)^F$.  The Hamiltonian we ultimately construct will be invariant under $\TAP$.

{\bf{Vector and axial vector symmetries:}}
Now let us discuss the form of the $U(1)_V$ and $U(1)_A$ symmetries.  We will write the corresponding generators first in the fermionic Villain setting, and then in the graded tensor product setting, but we will abuse notation slightly and use $Q_V$ and $Q_A$ for both settings.  This just means that $Q_V$ and $Q_A$ in the tensor product setting are defined by their images under $C$.  We define
\begin{align}
Q_V &= \frac{i}{2\pi}\sum_{r=0}^{N-1}\frac{d}{d\phi_r} = \frac{i}{2\pi}\sum_{r=0}^{N-1}\frac{d}{d\tphi_r}
\end{align}
which is simply the compact momentum of the $\phi$ rotors.  We also define
\begin{align}
Q_A &= 2\sum_{r=0}^{N-1}\left(\phi_{r+1} - \phi_r - n_{r,r+1}\right) \\ &= -2\sum_{r=0}^{N-1}n_{r,r+1} \text{    (on a ring)},
\end{align}
with the corresponding graded tensor product version being
\begin{align}
Q_A &= 2\sum_{r=0}^{N-1}\left(\tphi_{r+1} - \tphi_r - \frac{1}{2}\lr{2\tphi_{r+1} - 2\tphi_r}-\tn_{r,r+1}\right) \\ &= -\sum_{r=0}^{N-1}\left(\lr{2\tphi_{r+1} - 2\tphi_r}+ 2n_{r,r+1}\right) \text{    (on a ring)},
\end{align}
Note that we straightforwardly have
\begin{align}
\exp\left(\pi i Q_V\right) = (-1)^F
\end{align}
whereas
\begin{align} \label{eq:exp_pi_i_QA}
\exp\left(\pi i Q_A\right) = \exp\left(2\pi i \sum_{r=0}^{N-1} n_{r,r+1}\right).
\end{align}

{\bf{Dual constraints:}}
Now, our goal is to write a Hamiltonian for a Dirac fermion, for which we expect $\exp\left(\pi i Q_A\right) = (-1)^F$, which is different from eq. \ref{eq:exp_pi_i_QA}.  To implement $\exp\left(\pi i Q_A\right) = (-1)^F$, we will introduce constraints which are in a certain sense `dual' to the Villain conditions.  Like the Villain conditions, there will be $N$ independent, mutually commuting dual constraints, which (necessarily) commute with the Villain conditions, as well as the Hamiltonian we ultimately construct.  We could have in principle implemented these conditions as part of the definition of the Hilbert space, but this would have resulted in a Hilbert space that was not obviously equivalent to a graded tensor product of local graded Hilbert spaces.

We define
\begin{equation} \label{eq:defW}
W_{r,r+1} = i\gamma'_r \gamma_{r+1} \exp\left(2\pi i n_{r,r+1}\right)
\end{equation}
and by abuse of notation use the same letter to refer to the $C$ conjugated version in the graded tensor product Hilbert space:
\begin{equation}
W_{r,r+1} = i\gamma'_r \gamma_{r+1} \exp\left(2\pi i \tn_{r,r+1}\right) \exp \left(\pi i \lr{2\tphi_{r+1} - 2 \tphi_r}\right).
\end{equation}
We note that the $W_{r,r+1}$ have eigenvalues $\pm 1$, and that
\begin{align*}
\prod_{r=0}^{N-1} W_{r,r+1} = -(-1)^F \exp\left(2\pi i \sum_{r=0}^{N-1} n_{r,r+1}\right) = -(-1)^F \exp\left(\pi i Q_A\right)
\end{align*}
where the overall minus sign comes from a re-ordering of the Majorana fermions.  Since our goal is to have $\exp\left(\pi i Q_A\right) = (-1)^F$, we will impose $\prod_{r=0}^{N-1} W_{r,r+1}=-1$.  It turns out that we can do this in a way that preserves the anti-periodic translation symmetry $\TAP$, which acts on $W_{r,r+1}$ as follows:
\[ \label{eq:TWTinv}
\TAP W_{r,r+1} \TAP^{-1}=
\begin{cases}
W_{r+1,r+2}, & r = 1,\ldots, N-2,\\
-W_{r+1,r+2}, & r=0,N-1.
\end{cases}
\]
Indeed, we will energetically enforce the assignment $W_{r,r+1}=1$ for $r=0, \ldots, N-2$ and $W_{N-1,0} = -1$; this is invariant under $\TAP$, as can be seen from above.

\section{Hamiltonian} \label{sec: hamiltonian}
Our Hamiltonian will be:
\begin{align}
H &= \frac{1}{2}\sum_{r=0}^{N-1}\left(-U_0 \frac{d^2}{d\phi_r^2} + J_0 \left(\phi_{r+1}-\phi_r- n_{r,r+1}\right)^2\right) \notag \\ &+V\left(W_{0,N-1}-\sum_{r=0}^{N-2}W_{r,r+1}\right) \label{eq: fermionic Villain Hamiltonian}
\end{align}
This realizes a general interacting fermionic system in the Luttinger liquid universality class; we will see that the specific choice $U_0/J_0 = 1/({16 \pi^2})$ corresponds to the free Dirac fixed point and $V$ is a large energy scale.  Note that the harmonic terms on the first line of the Hamiltonian commute with all $W_{r,r+1}$, so the latter may indeed be thought of as mutually commuting constraints.  Although the Hamiltonian appears purely bosonic, the fermions enter through the constraints $W_{r,r+1}$ and the fermionic Villain constraint.

This Hamiltonian is exactly solvable for essentially the same reason as that in ref. \cite{cheng_lieb-schultz-mattis_2023}: the field $n$ and the fermions have no dynamics, and gauge-fixing the fermionic Villain constraint reduces the problem to coupled simple harmonic oscillators.  Let us demonstrate this solution in three sectors: $Q_A=0$, even non-zero $Q_A$, and odd $Q_A$.  We will take odd $N$ for convenience throughout.

{\bf{Zero axial charge ($Q_A=0$)}:}
By definition of $Q_A$ this implies that $\sum_{r=0}^{N-1} n_{r,r+1} = 0$.  We also have $(-1)^F = \exp(\pi i Q_A)=1$.  We will gauge fix $n_{r,r+1}=0$ for all $r$, which, using the $W_{r,r+1}$ constraints forces $i\gamma'_r\gamma_{r+1}=1$ for $r=0,\ldots, N-2$ and $i\gamma'_{N-1} \gamma_0 = -1$.  We will solve the Hamiltonian in this gauge fixed sector, ignoring the fermionic Villain condition for the moment, and then map the solution to the other sectors by conjugating by the fermionic Villain condition, which restores invariance under it.  In this gauge-fixed sector, the Hamiltonian is simply the following set of (bosonic) coupled harmonic oscillators:
\begin{align*}
H_{\text{eff}} &= \frac{1}{2}\sum_{r=0}^{N-1}\left(-U_0 \frac{d^2}{d\phi_r^2} + J_0 \left(\phi_{r+1}-\phi_r\right)^2\right)
\end{align*}
with the wavefunction being invariant under simultaneous shifts $\phi_r \rightarrow \phi_r + \frac{1}{2}$.

To solve, we expand in plane-wave modes (recall we assume odd $N$ throughout):
\begin{align*}
\phi_r = \sqrt{\frac{2}{N}}\sum_{k=1}^{(N-1)/2}\left(A_k \cos\frac{2\pi k r}{N} + B_k \sin \frac{2\pi k r}{N}\right) + \frac{1}{\sqrt{N}}A_0
\end{align*}
where the expansion coefficients $A_k, B_k, A_0$ should be viewed as operators.  The inverse relation is
\begin{align*}
A_k &= \sqrt{\frac{2}{N}} \sum_{r=0}^{N-1}\phi_r \cos \frac{2\pi k r}{N} \text{ for }k= 1,\ldots,(N-1)/2 \\
B_k &= \sqrt{\frac{2}{N}} \sum_{r=0}^{N-1}\phi_r \sin \frac{2\pi k r}{N} \text{ for }k= 1,\ldots,(N-1)/2 \\
A_0 &= \frac{1}{\sqrt{N}} \sum_{r=0}^{N-1} \phi_r
\end{align*}
Since the transformation between the $\phi_r$ and the $A_k, B_k, A_0$ is orthogonal, we can readily write the Hamiltonian in the plane-wave coordinates, obtaining

\begin{align}\label{eq:Hamiltonian_k_space}
H_{\text{eff}} &= -\frac{1}{2}U_0\left(\frac{d^2}{d{A_0}^2} + \sum_{k=1}^{(N-1)/2}\left(\frac{d^2}{d{A_k}^2}+ \frac{d^2}{d{B_k}^2} \right) \right) \notag \\ &+J_0\sum_{k=1}^{(N-1)/2}\left(1-\cos \frac{2\pi k}{N}\right)(A_k^2+B_k^2)
\end{align}
In particular, the dispersion of the $k \neq 0$ normal modes is (setting $\hbar=1$):
\begin{align}
\omega_k = \sqrt{2U_0 J_0\left(1-\cos \frac{2\pi k}{N}\right)} \label{eq: Villain spectrum}
\end{align}
which implies that the group velocity of long wavelength excitations is $\frac{d\omega_k}{dk}|_{k=0} = a\sqrt{U_0 J_0}$.  There is also a zero mode which has to be treated separately.  The wavefunction is invariant under $\phi_r \rightarrow \phi_r + \frac{1}{2}$ for all $r$, which sends $A_0 \rightarrow A_0 + \frac{\sqrt{N}}{2}$.  The wavefunctions for the zero mode have the form
\begin{align*}
\exp\left(2 \pi i m \frac{A_0}{\sqrt{N}/2}\right)
\end{align*}
where $m\in \Z$.  The associated energy is $8\pi^2 m^2 U_0/N$.

This sector ($Q_A = 0$) describes the sector of the field theory of the $c=1$ compact boson with even fermion parity under bosonization.  Specifically, this is the sector with equal left and right moving charges, with the harmonic excitations of the oscillators describing the particle-hole excitations above the Fermi sea.  This is consistent with the fact that $(-1)^F=1$ for this sector in our lattice model.

{\bf{Even non-zero axial charge $Q_A = 2\bar{n}, \bar{n} \in \Z$}:}
We again gauge fix the Majorana fermions as in the $Q_A$ case, and furthermore gauge fix $n_{r,r+1}$ such that $n_{N-1,0} = {\bar{n}}$.  We define new variables $\phi'_r$:
\begin{align*}
\phi_r = \phi'_r - \frac{r\bar{n}}{N}
\end{align*}
Then for all $r=0,\ldots, N-1$
\begin{align*}
\phi_{r+1}-\phi_r - n_{r,r+1} = \phi'_{r+1}-\phi'_r - \frac{\bar{n}}{N}
\end{align*}
so we obtain an effective Hamiltonian, in the space where the dual constraints are imposed, of the form
\begin{align*}
H_{\text{eff}} &= \frac{1}{2}\sum_{r=0}^{N-1}\left(-U_0 \frac{d^2}{d{\phi'_r}^2} + J_0 \left(\phi'_{r+1}-\phi'_r\right)^2\right) + \frac{1}{2}J_0\frac{{\bar{n}}^2}{N}
\end{align*}
This gives the same exact effective Hamiltonian as in the $Q_A=0$ case, including the same zero mode, with the same periodicity (because $(-1)^F=1$) and spectrum.  Physically, this is the sector with different left and right moving charges, but where the difference is even.

{\bf{Odd non-zero axial charge $Q_A = 2\bar{n}, \bar{n} \in \Z+\frac{1}{2}$:}}
We perform the same analysis as before, except now, when we gauge fix $n_{N-1,0}={\bar{n}}$, we must fix $i\gamma'_{N-1} \gamma_0=1$ to satisfy $W_{N-1,0}=-1$, so that all $i\gamma'_r \gamma_{r+1}=1$.  This gives $(-1)^F=-1$, so all of these states have odd fermion parity.  The solution is again the same as before, except the zero mode has anti-periodic boundary conditions in field space (not real space), as follows from the fact that $\exp(\pi i Q_V) = (-1)^F = -1$.
Thus the zero mode wavefunctions are again
$\exp\left(2 \pi i m \frac{A_0}{\sqrt{N}/2}\right)$ but now with $m \in \Z+\frac{1}{2}$.

This sector contains all of the odd fermion parity excitations, including a single left-mover and a single right-mover in the case of the free Dirac fixed point.  Let us calculate the ratio $U_0/J_0$ for this free Dirac fixed point.  In fact, comparing the energy of a single chiral mover to a pair of counter-propagating fermions allows us to fix this ratio.  Indeed, for free fermions and large $N$ we expect this ratio to reproduce the ratio of scaling dimensions of the corresponding fields $\psi_L$ and $\psi_L \psi_R$.  The former has dimension $1/2$ and the latter has dimension $1$ at the free Dirac fixed point.  We thus obtain
\begin{align*}
\frac{\frac{2\pi^2U_0}{N}+\frac{1}{2}\frac{J_0/4}{N}} {\frac{8\pi^2U_0}{N}}= \frac{1}{2}
\end{align*}
from which we obtain $U_0/J_0 = \frac{1}{16\pi^2}$.  
This value of $U_0/J_0$ will be crucial in ensuring that the operators ${\mathcal{F}}^\dagger_{L/R}$ we write down in the next section are chiral - i.e. when acting on the ground state, they create only left movers or only right movers.  This is true at the free Dirac fixed point, but not in the generic Luttinger liquid, as follows from the fact that in a generic Luttinger liquid the holomorphic and anti-holomorphic scaling dimensions for the corresponding primary field are both non-zero.

In summary, our Hamiltonian is effectively a free boson at low energies, but the states with half-odd winding (i.e. odd $Q_A$) are fermion-parity odd.  This is then, at low energies, precisely a Dirac fermion when $U_0/J_0$ is tuned to $\frac{1}{16\pi^2}$.

\section{Fermion parity odd operators} \label{sec: fermion parity odd operators}


A natural question is, how to construct fermion parity odd operators that act within the low energy space of our Hamiltonian, whose low energy theory is a Dirac fermion.  While fermion parity even operators that correspond to particle-hole excitations are easy to write down, fermion parity odd operators are more subtle.  We will define an operator that pulls in a single chiral fermion from the Dirac sea, by shifting all the energy levels of that dispersion up by one (while leaving the counter-propagating dispersion alone).  The operator is
\begin{align}
{\mathcal{F}}^\dagger_L&=AB \label{eq: delocalized left} \\
{\mathcal{F}}^\dagger_R&=A^{-1}B \label{eq: delocalized right}
\end{align}
where
\begin{align}
A = \prod_{r=0}^{N-1}\exp\left(-\frac{r}{2N}\frac{d}{d\phi_r}\right) \exp\left(-i\chi_{N-1,0}\right)
\end{align}
and
\begin{align}
B = \prod_{r=0}^{N-1}\exp\left(\frac{2\pi i r}{N}(\phi_{r+1}-\phi_r - n_{r,r+1})\right) \exp\left(-2\pi i \phi_0\right)\gamma_0.
\end{align}
We show explicitly in Appendix \ref{sec: solving model} the operators ${\mathcal{F}}^\dagger_{L/R}$ operate only on the zero modes of the bosonic field.  By construction $A$ increments $Q_A$ by $1$ and $B$ increments $Q_V$ by $1$.  So the combination ${\mathcal{F}}^\dagger_L=AB$ increments $Q_V + Q_A$ by $2$ while keeping $Q_V-Q_A$ constant.  Likewise ${\mathcal{F}}^\dagger_R = A^{-1}B$ keeps $Q_V + Q_A$ fixed, and increments $Q_V - Q_A$ by $2$.  Identifying 
\begin{align*}
Q_L &= \frac{1}{2}(Q_V + Q_A)\\
Q_R &= \frac{1}{2}(Q_V - Q_A)
\end{align*}
we see that ${\mathcal{F}}^\dagger_{L/R}$ raise the corresponding chiral charges by $1$.

It is important to note that ${\mathcal{F}}^\dagger_{L/R}$ do not obey the usual Fock space relations of a single fermionic mode.  They are in fact unitary operators whose spectrum contains all complex numbers of unit modulus.  Indeed, because they only act on the zero modes, it is trivial to check that they send eigenstates of our Hamiltonian to other eigenstates, i.e. do not create any particle-hole excitations.  The action of ${\mathcal{F}}^\dagger_{L}$ is to add a unit of left-moving charge by pulling it in from the Dirac sea; arbitrary amounts of charge can be added by shifting the spectrum arbitrarily far.

Explicit computation shows that, using the fermionic Villain condition, we obtain
\begin{align*}
\TAP A\, \TAP^{-1}=A\,(i\gamma_0\gamma_0')\prod_{r=0}^N \exp\left(-\frac{1}{2N}\frac{d}{d\phi_r}\right)
\end{align*}
and
\begin{align*}
\TAP B \,\TAP^{-1} = B \gamma_0\exp\left(-2\pi i n_{0,1} \right) \gamma_1 \prod_{r=0}^{N-1} \exp\left(\frac{2\pi i}{N}n_{r,r+1}\right)
\end{align*}
which results in
\begin{align*}
\TAP AB \,\TAP^{-1} &= \TAP {\mathcal{F}}_L^\dagger \TAP^{-1} \\ 
&= \exp\left(-\frac{2\pi i}{2N}\right) AB \prod_{r=0}^{N-1} \exp\left(-\frac{1}{2N}\frac{d}{d\phi_r}\right) \\ &\cdot \prod_{r=0}^{N-1}\exp\left(\frac{2\pi i}{N} n_{r,r+1}\right) \cdot i\gamma_0'\gamma_1 \exp(-2\pi i n_{0,1}) \\ &= \exp\left(-\frac{2\pi i}{2N}\right) {\mathcal{F}}_L^\dagger 
\exp\left(\frac{\pi i}{N}(Q_V - Q_A)\right)
\end{align*}
using $W_{0,1}=1$.  In particular, when acting on the left-moving chiral sector only, which has $Q_R = \frac{1}{2}(Q_V - Q_A) = 0$, we have
\begin{align*}
\TAP {\mathcal{F}}_L^\dagger \TAP^{-1} = \exp\left(-\frac{2\pi i}{2N}\right) {\mathcal{F}}_L^\dagger 
\end{align*}
which is consistent with the anti-periodic boundary conditions and the odd fermion parity nature of ${\mathcal{F}}_L^\dagger$.

The fact that ${\mathcal{F}}_L^\dagger$ acts only by shifting all the left moving fermionic modes can be argued from the fact that, because it acts only on the bosonic zero modes, it commutes with all the non-zero momentum oscillator modes.  These create particle-hole excitations, of either chirality, and any configuration of particles and holes can be created by appropriate combinations of these modes acting on the ground state.  Now, when one of these non-zero oscillator modes gets conjugated by ${\mathcal{F}}_L^\dagger$, it simply creates the same configuration of particles and holes, but relative to the new, shifted Fermi point.  This is just the statement that ${\mathcal{F}}_L^\dagger$ shifts any configuration of particles and holes up by $1$.

{\bf{Fermion parity odd operators of finite support:}} A natural question is whether the non-local operators ${\mathcal{F}}_{L/R}^\dagger$ can be made local.  For example, imagine that $N$ is very large - effectively infinite, so our system is on an infinite line - and we want to localize ${\mathcal{F}}_{L}^\dagger$ to an interval $I$ consisting of $|I|$ lattice points.  The natural generalization of ${\mathcal{F}}_{L}^\dagger$ is an operator ${{\mathcal{F}}_{L}^I}^\dagger$ that inserts winding and charge only in $I$.  For instance, we can let $f(r)$ be a function that is $0$ to the left of $I$, $\frac{1}{2}$ to the right of $I$, and interpolates between those values within $I$.  Similarly, let $g(r)$ be a function that is $0$ to the left of $I$, $1$ to the right of $I$, and interpolates between these two in some (potentially different) way.  Then we can define
\begin{align}
A^I = \prod_{r=0}^{N-1}\exp\left(-f(r)\frac{d}{d\phi_r}\right) \exp\left(-i\chi_{R,R+1}\right) 
\end{align}
and
\begin{align}
B^I = \prod_{r=0}^{N-1}\exp\left(2\pi i g(r)(\phi_{r+1}-\phi_r - n_{r,r+1})\right) \exp\left(-2\pi i \phi_R\right)\gamma_R.
\end{align}
where $R$ is the right endpoint of $I$.  There is implicit dependence in $A^I$ and $B^I$ on $f$ and $g$ which we are suppressing.  From these operators, we can define
\begin{align}
{\mathcal{F}^I}^\dagger_L&=A^I B^I \label{eq: localized left} \\
{\mathcal{F}^I}^\dagger_R&=\left(A^I\right)^{-1} B^I \label{eq: localized right}
\end{align}
These still have well defined $Q_L$ and $Q_R$ charges, as before.  Also, in the large $|I|$ limit, for any $f$ we can find a $g$ such that the resulting operators are approximately chiral.  This just comes down to appropriately setting the initial momentum given the initial displacement of a kink to make it a chiral mover.  However, as we show in Appendix \ref{sec: solving model}, for any finite $|I|$ the corresponding operators ${\mathcal{F}^I}^\dagger_{L/R}$ will involve more than just the zero modes of the bosonic field. In fact, the discrete Fourier transforms of $f$ and $g$ cannot be completely localized in a finite region of reciprocal space, due to the fact that their gradients are identically zero outside of the real space interval $I$. Explicit computation also shows that it is at the free fermion radius ($U_0/J_0 = 1/16\pi^2$) where a localized fermion ${\mathcal{F}^I}^\dagger_{\nu}$ creates the fewest amount of particle-hole pairs in the $\nu$ branch at the cost of creating many pairs in the opposite branch.


\section{Jordan-Wigner transform and lattice bosonization with open boundary conditions}

Motivated by the continuum bosonization, we can also define a lattice level Jordan-Wigner-like transformation from the fermionic tensor product Hilbert space to a bosonic tensor product Hilbert space, which, when implemented on our Hamiltonian, performs the field theoretic bosonization.  This is easiest to do with open boundary conditions, so in this section we imagine that the lattice sites $0, \ldots, N-1$ are on an interval.  We define the bosonic system to be built from rotors $\tphi_r' \in \R / \Z$ and integer valued variables $\tn'_{r,r+1} \in \Z / 2$.  We let $\exp(i\tchi'_{r,r+1})$ denote the dual of $\tn'_{r,r+1}$.  We then define the Jordan Wigner map on the fermion parity even subalgebra as
\[
\begin{array}{rcl}
C_{\text{JW}}:\qquad
\exp(4\pi i \tphi_r) &\longmapsto& \exp(4\pi i \tphi'_r),\\
\tn_{r,r+1} &\longmapsto& \tn'_{r,r+1},\\
\frac{d}{d\tphi_r} &\longmapsto& \frac{d}{d\tphi'_r},\\
\exp(i\tchi_{r,r+1}) &\longmapsto& \exp(i\tchi'_{r,r+1}),\\
i\gamma'_r \gamma_{r+1}\exp(2\pi i (\tphi_{r+1}-\tphi_r)) &\longmapsto& \exp(2\pi i (\tphi_{r+1}-\tphi_r)).
\end{array}
\]
whereas on the odd operator $\gamma_r \exp\left(2\pi i \tphi_r\right)$ we define its action to be 
\[
\begin{array}{rcl}
C_{\text{JW}}:\qquad
\gamma_r \exp\left(2\pi i \tphi_r\right) &\longmapsto& \prod_{r'<r} \exp\left(\frac{1}{2}\frac{d}{d\tphi'_r}\right) \exp\left(2\pi i \tphi_r\right)
\end{array}
\]

\section{Scattering analysis of irrelevant interactions}
Throughout this section we will work with the free Dirac fixed point.  As found earlier, the spectrum of the Hamiltonian is non-linear, whereas standard bosonization at the free fermion duality point assumes that the spectrum is linear. Thus one could ask: what are the signatures of this non-linearity, and how do they manifest themselves in the low energy fermionic system? We probe this in the following way. Say that, after diagonlizing the Villain Hamiltonian, one has a set of bosonic creation (annihilation) operators $B^\dag_n$ ($B_n$) where these operators correspond to creating (annihilating) an excitation at momentum $2\pi n/N$, with $n \in \mathbb{Z}$ and $N$ being the total number of lattice sites. Also say that $\ket{\tilde\Omega}$ is the ground state in the Villain model defined through the relation $B_n \ket{\tilde\Omega} = 0$ for all $n$. If the spectrum of the model was linear, then the states $\left(B_1^\dag \right)^2\ket{\tilde\Omega}$ and $B_2^\dag \ket{\tilde\Omega}$ would have the same energy. However, the non-linearity causes a splitting between the energies of these states.  This turns out to have an important consequence for the scattering of fermions, as we explain below.

In the setting of field-theoretic bosonization, one can construct the analogous states using the bosonization dictionary. Using the conventions of ref. \cite{delft_bosonization_1998}, where the system lives on a continuous space with length $L$, the fermionic annhilation operator in position and momentum space are 
\begin{align}
    \psi_{\nu}(x)  &= \left(\frac{2\pi}{L}\right)^{1/2}\sum_{n\in\mathbb{Z}} e^{-\nu \cdot ik_nx} c_{n,\nu} \\
    c_{n,\nu} &= \left(\frac{1}{2\pi L}\right)^{1/2} \int_{-L/2}^{L/2} dx \; e^{\nu \cdot ik_n x} \psi_{\nu}(x)
\end{align}
where $k_n = 2\pi n/L$, $n\in \mathbb{Z}$, and $\nu = \{L,R\} = \{+1, -1\}$ labels the chirality of the fields. The bosonic creation operator corresponding to the ladder operator for the free boson Hamiltonian is

\begin{equation}
    b_{n,\nu}^\dag \equiv \frac{i}{\sqrt{n}} \sum_{n'\in\mathbb{Z}} c_{n+n', \nu}^\dag c_{n', \nu},
\end{equation}
with $n\in \mathbb{Z}^+$. For the rest of this section, we will work with left handed fermions ($\nu = +1$) and suppress the chiral label. In this framework, the states analogous the Villain model states introduced earlier are 

\begin{align}
    \left(b_1^\dag \right)^2 \ket{\Omega} &= (c_1^\dag c_{-1} - c_2^\dag c_0) \ket{\Omega} \\
    b^\dag_{2}\ket{\Omega} &= \frac{i}{\sqrt{2}} (c_2^\dag c_0 + c^\dag_1 c_{-1})\ket{\Omega}.
\end{align}

If we are working directly at the duality point, the free boson Hamiltonian will not mix these two states. So in order to generate mixing, one must introduce an interaction that couples these states. Additionally, to stay at the duality point under renormalization group flow, the interaction must be irrelevant. If we want to write the lowest order irrelevant interaction using fermions from a single chiral branch, then (up to total derivatives and anticommutation) we have a unique choice of interaction:

\begin{equation}
    H_I = \lambda \int dx \; \psi^\dag \partial_x \psi^\dag \partial_x \psi \psi.
\end{equation}
Now we consider the states

\begin{align}
    \ket{\psi_+} \equiv c^\dag_1 c_{-1}\ket{\Omega} &= \frac{1}{\sqrt{2}i}\left(b_2^\dag + \frac{i}{\sqrt{2}}(b_1^\dag)^2 \right)\ket{\Omega} \label{eq: defn of psi plus} \\
    \ket{\psi_-} \equiv c^\dag_2 c_0\ket{\Omega} &= \frac{1}{\sqrt{2}i}\left(b_2^\dag - \frac{i}{\sqrt{2}} (b_1^\dag)^2 \right)\ket{\Omega} \label{eq: defn of psi minus}. 
\end{align}
One can straightforwardly check that these states are orthonormal. Thus if we start with an initial state $\ket{\psi_-}$, under the free Hamiltonian $H_0$ there will be no overlap with $\ket{\psi_+}$. However, if we introduce $H_I$ this will no longer be the case. In particular, after a short amount of time we have 

\begin{equation}
    A(t) \equiv \bra{\psi_+} e^{-i(H_0 + H_I)t} \ket{\psi_-} \sim -i \bra{\psi_+}H_I \ket{\psi_-}t.
\end{equation}
Our goal will then be to calculate $A(t)$, which will be a function of the coupling constant $\lambda$, and the analogous quantity to $A(t)$ in the Villain Hilbert space, compare the answers in order to make a prediction for the value of $\lambda$. 

\subsection{Mixing in the Fermionic Field Theory}
Writing $H_I$ in momentum space

\begin{equation*}
    H_I = \lambda \frac{(2\pi)^2}{L} \sum_{\alpha,\beta,\gamma,\delta \in \mathbb{Z}} \delta_{\alpha + \beta, \gamma + \delta} c^\dag_{\alpha} c^\dag_{\beta} c_{\gamma} c_{\delta} (ik_\beta)(-ik_\gamma),
\end{equation*}
we have for the mixing matrix element

\begin{align}
    \bra{\psi_+}H_I\ket{\psi_-} &= \lambda \frac{(2\pi)^2}{L} \sum_{\alpha,\beta,\gamma,\delta} \bra{\Omega}c^\dag_{-1} c_1 c_\alpha^\dag c_\beta^\dag c_\gamma c_\delta c^\dag_2 c_0\ket{\Omega} \delta_{\alpha + \beta, \gamma + \delta} (ik_\beta)(-ik_\gamma). \label{eq: mixing matrix element}
\end{align}
Considering the amplitude in eq. \ref{eq: mixing matrix element}, by Wick's theorem, there are four possible pairings of momenta that contribute to this:

\begin{table}[H]
    \centering
    \begin{tabular}{c|c|c|c}
        $\alpha$ & $\beta$ & $\gamma$ & $\delta$ \\
        \hline
        0 & 1 & -1 & 2 \\
        0 & 1 & 2 & -1 \\
        1 & 0 & -1 & 2 \\
        1 & 0 & 2 & -1
    \end{tabular}
    \label{tab:placeholder}
\end{table}
However, notice that since $k_\beta = 0$ for the third and fourth pairings, by comparing to eq. \ref{eq: mixing matrix element}, we can see these terms will vanish. Therefore, there are two non-zero terms in eq. \ref{eq: mixing matrix element}, corresponding to 

\begin{align}
    \bra{\Omega} c^\dag_{-1} c_1 c_0^\dag c_1^\dag c_{-1} c_2 c^\dag_2 c_0 \ket{\Omega} (k_1 \cdot k_{-1}) &= -\left(\frac{2\pi}{L} \right)^2 \\
    \bra{\Omega}c^\dag_{-1} c_1 c_0^\dag c_1^\dag c_2 c_{-1} c^\dag_2 c_0\ket{\Omega} (k_1 \cdot k_2) &= -2\left(\frac{2\pi}{L} \right)^2,
\end{align}
which leads to

\begin{equation}
    A(t) = 3\lambda \cdot it\frac{(2\pi)^4}{L^3}. \label{eq: mixing amplitude fermionic FT}
\end{equation}

\subsection{Mixing in the Villain Model}
To calculate the analogous quantity to $A(t)$, we consider the states

\begin{equation}
    \ket{\tilde\psi_\pm} = \frac{1}{\sqrt{2}i}\left(B_2^\dag \pm \frac{i}{\sqrt{2}}\left(B_1^\dag\right)^2\right) \ket{\tilde\Omega},
\end{equation}
where we introduce the tilde to emphasize these are analogous to but not the same as the states in the field theory. The mixing matrix element for short times is

\begin{equation}
    \bra{\tilde\psi_+}e^{-iH_{Vil}t}\ket{\tilde\psi_-} = \frac{1}{2}\left(e^{-i\omega_2t} - e^{-2i\omega_1t}\right) \sim \frac{it}{2}(2\omega_1 - \omega_2),
\end{equation}
where $\omega_1$ and $\omega_2$ are the energies of the states $B_1^\dag\ket{\tilde\Omega}$ and $B_2^\dag\ket{\tilde\Omega}$, respectively, which were found in eq. \ref{eq: Villain spectrum}. As the Villain model has no dimensionful size, in order to have a proper continuum limit one should identify $N = L/a$, where $a$ is the lattice spacing. Thus, for large $N$ we have

\begin{equation}
    2\omega_1 - \omega_2 = 2\sqrt{J_0U_0}\left(\frac{a\pi}{L} \right)^3,
\end{equation}
leading to

\begin{equation}
    \bra{\tilde\psi_+}e^{-iH_{Vil}t}\ket{\tilde\psi_-} \sim it \sqrt{J_0 U_0} \left(\frac{a\pi}{L} \right)^3. 
\end{equation}
Comparing this expression to eq. \ref{eq: mixing amplitude fermionic FT} we get a prediction for $\lambda$:

\begin{equation}
    \lambda = \frac{a^3}{48\pi}\sqrt{J_0U_0}.
\end{equation}

\section{Discussion}
In this work we constructed a lattice Hamiltonian for a $1+1$d Dirac fermion, with exact $U(1)_V$ and $U(1)_A$ symmetries.  Due to the mixed anomaly between them, one of these - in our case, $U(1)_A$ - is necessarily realized not-on-site.  Our lattice Hilbert space can be viewed either as that of a fermionic Villain model or a graded tensor product of $\Z_2$-graded local Hilbert spaces.  Importantly, these are intrinsically fermionic Hilbert spaces.  Our Hamiltonian can be viewed as a fermionized version of that in ref. \cite{cheng_lieb-schultz-mattis_2023}, which makes the appropriate kinks of the low energy compact boson into odd fermion parity excitations.  We also constructed odd fermionic operators, which can be viewed as lattice realizations of chiral fermions.  These are non-local, but can be made approximately local when restricted to a large interval $I$, with the degree of approximation controlled by $e^{-c|I|}$, where $c$ is a positive constant.

One natural avenue for further investigation is to use this work to construct a fermionic lattice Hamiltonian version of the $3450$ model \cite{thorngren_chiral_2026, seifnashri_exactly_2026, berkowitz_exact_2024} (in contrast to say Euclidean lattice methods \cite{gorantla_modified_2021,demarco_lattice_2023}, or Hamiltonian approaches that require analyzing a strongly interacting system \cite{zeng_symmetric_2022,wang_solution_2019}) with the anomaly-free $U(1)$ symmetry realized onsite, or at least in a way that makes it readily gaugable. 

{\bf{Acknowledgments:}} LF would like to acknowledge illuminating discussions with Ryan Thorngren.  LF is supported by NSF DMR-2300172.

\bibliographystyle{unsrt}
\bibliography{lattice_bosonization4}

\appendix

\section{Particle-Hole Creation} \label{sec: solving model}
In section \ref{sec: hamiltonian}, we solved the fermionic Villain model by choosing a gauge where the $n$ fields are zero in the bulk of the system. Here, we will solve the model without having to make this choice. This allows for an explicit expression of the mode operators of the fermionic Villain model (which, under the familiar bosonization map correspond to the operators that create particle-hole pairs), and subsequently by checking the commutator of our proposed delocalized and localized single fermion operators with these mode operators we can see whether they can genuinely be thought of as single particle operators. This method to solve the model closely follows that of solving the bosonic Villain model presented in ref. \cite{seifnashri_exactly_2026}. \\

First we introduce the following staggered operator $\Phi_i$ for $i = 0,...,2N-1$:

\begin{align}
    \Phi_{2j} &= \sqrt{U_0}\Pi_j \\
    \Phi_{2j+1} &= \sqrt{J_0}D\phi_{j,j+1}
\end{align}
where we've also introduced the discrete gauge invariant derivative

\begin{equation}
    D\phi_{j,j+1} \equiv \phi_{j+1} - \phi_j - n_{j,j+1}.
\end{equation}
We also define the Fourier transform of this field $M_k$ for $k = 0,...,2N-1$:

\begin{align}
    M_k &\equiv \frac{1}{\sqrt{2N}}\sum_{j=0}^{2N-1} e^{\frac{2\pi i}{2N}jk} \Phi_j \\
    &= \frac{1}{\sqrt{2N}}\sum_{j=0}^{N-1} e^{\frac{2\pi i}{N}jk}\left(\sqrt{U_0}\Pi_j + e^{\frac{\pi i}{N}k}\sqrt{J_0}D\phi_{j,j+1} \right)
\end{align}
which due to the Hermiticity of $\phi_j$ and $\Pi_j$ means $M_k^\dag = M_{-k}$. One can also show that the operator algebra of $\Phi_j$ and $M_k$ is the following:

\begin{align}
    &[\Phi_j, \Phi_{j'}] = i\sqrt{U_0J_0} \cdot (\delta_{j',j+1} - \delta_{j',j-1}) \\
    &[M_k,M_{k'}^\dag] = \omega_k \cdot \delta_{k,k'} \label{eq: M algebra}
\end{align}
where $\omega_k = 2\sqrt{U_0J_0}\sin(\frac{\pi k}{N})$. We lastly define $B_k = M_k/\sqrt{\omega_k}$ where $k=1,...,N-1$. By inspecting eq. \ref{eq: M algebra}, $B_k^\dag$ and $B_k$ satisfy the bosonic creation and annihilation operator algebra. We can write the fermionic Villain Hamiltonian (eq. \ref{eq: fermionic Villain Hamiltonian}) in terms of these operators in the following ways:

\begin{align}
    H &= \frac{1}{2}\sum_{j=0}^{N-1}\left(U_0 \Pi_j^2 + J_0 (D\phi_{j,j+1})^2\right) + H_{\text{int}} \\
    &= \frac{1}{2}\sum_{j=0}^{2N-1}\Phi_j^2 + H_{\text{int}} \\
    &= \frac{1}{2}\sum_{k=0}^{2N-1}M_k^\dag M_k + H_{\text{int}} \\
    &= \frac{1}{2} \left(M_0^\dag M_0 + M_N^\dag M_N\right) + \sum_{k=1}^{N-1} \omega_k\left(B_k^\dag B_k + \frac{1}{2} \right) + H_{\text{int}}
\end{align}
where we called the dual constraint term

\begin{equation}
    H_{\text{int}} = V\left(W_{0,N-1}-\sum_{j=0}^{N-2}W_{j,j+1}\right).
\end{equation}
Note that the $M^\dag_0 M_0$ and $M^\dag_N M_N$ terms are quadratic in $Q_V$ and $Q_A$ as

\begin{align}
    M_0 &= \frac{1}{\sqrt{2N}} \left(-2\pi\sqrt{U_0}Q_V + \frac{\sqrt{J_0}}{2}Q_A \right) \\
    M_N &= \frac{1}{\sqrt{2N}} \left(-2\pi\sqrt{U_0}Q_V - \frac{\sqrt{J_0}}{2}Q_A \right).
\end{align}

Now we will check the claims made in section \ref{sec: fermion parity odd operators}: first that the delocalized fermion operators (eqs. \ref{eq: delocalized left} and \ref{eq: delocalized right}) commute with the particle-hole operators; second that if the size of the non-trivial region, $I$, of the localized fermion operators (eqs. \ref{eq: localized left} and \ref{eq: localized right}) is large compared to the lattice spacing but small compared to the size of the system, then the localized fermions will have an exponentially small commutator with the particle-hole operators.

Let's first check the commutator of the delocalized fermions with the particle-hole operators. First note the following:

\begin{align*}
    &[\mathcal{F}^\dag_{L/R},\Pi_j] = \frac{2\pi}{N} \cdot \mathcal{F}^\dag_{L/R} \\
    &[\mathcal{F}^\dag_{L/R},\phi_j] = \mp \frac{j}{2N} \cdot \mathcal{F}^\dag_{L/R} \\
    &[\mathcal{F}^\dag_{L/R}, n_{j,j+1}] = \pm \frac{\delta_{j,N-1}}{2} \cdot \mathcal{F}^\dag_{L/R}
\end{align*}
Therefore we have 

\begin{align}
    [\mathcal{F}^\dag_{L/R}, M_k] &= \frac{1}{\sqrt{2N}}\sum_{j=0}^{N-1} e^{\frac{2\pi i}{N}jk}\left(\sqrt{U_0}[\mathcal{F}^\dag_{L/R}, \Pi_j] + e^{\frac{\pi i}{N}k}\sqrt{J_0}[\mathcal{F}^\dag_{L/R}, D\phi_{j,j+1}] \right) \notag \\
    &= (\delta_{k,0} + \delta_{k,N})  \cdot \frac{1}{\sqrt{2N}} \left(2\pi\sqrt{U_0} \mp \frac{\sqrt{J_0}}{2}e^{\frac{\pi i}{N}k} \right) \mathcal{F}^\dag_{L/R},
\end{align}
where the dependence on only $k=0$ and $N$ comes from the fact that the delocalized fermions can be thought of as being constructed in order to have the argument of the sum not depend on the spacial position. This tells us that 

\begin{equation}
    [\mathcal{F}^\dag_{L/R}, B_k] = 0,
\end{equation}
showing that the delocalized fermions do not change the particle-hole content of a state.

Now we check a similar claim for the localized fermions. First we note analogous properties as above:

\begin{align*}
    &[{\mathcal{F}^I}^\dagger_{L/R},\Pi_j] = 2\pi dg_{j-1,j} \cdot {\mathcal{F}^I}^\dagger_{L/R} \\
    &[{\mathcal{F}^I}^\dagger_{L/R},\phi_j] = \mp f(j) \cdot {\mathcal{F}^I}^\dagger_{L/R} \\
    &[{\mathcal{F}^I}^\dagger_{L/R}, n_{j,j+1}] = \pm \frac{\delta_{j,N-1}}{2} \cdot {\mathcal{F}^I}^\dagger_{L/R}
\end{align*}
Thus

\begin{align}
    [{\mathcal{F}^I}^\dagger_{L/R}, M_k] &= \frac{1}{\sqrt{2N}}\sum_{j=0}^{N-1} e^{\frac{2\pi i}{N}jk}\left(\sqrt{U_0}[{\mathcal{F}^I}^\dagger_{L/R}, \Pi_j] + e^{\frac{\pi i}{N}k}\sqrt{J_0}[{\mathcal{F}^I}^\dagger_{L/R}, D\phi_{j,j+1}] \right) \notag \\
    &= \frac{1}{\sqrt{2N}}\sum_{j\in I} e^{\frac{2\pi i}{N}jk}\left(2\pi\sqrt{U_0}dg_{j-1,j} \mp e^{\frac{\pi i}{N}k}\sqrt{J_0} df_{j,j+1} \right) {\mathcal{F}^I}^\dagger_{L/R} \label{eq: localized commutator},
\end{align}
i.e. the amount of particle-hole content that ${\mathcal{F}^I}^\dagger_{L/R}$ create is closely related to the discrete Fourier transform of $df$ and $dg$ in $I$. One could choose $f$ and $g$ to be a linear interpolation, which would mean that the nonzero values of the sum in eq. \ref{eq: localized commutator} would be at the edges of $I$ with a size going as $\sim 1/|I|$. Another natural choice would be having $df$ and $dg$ go as Gaussians.  Note that in this case the profiles are not strictly localized inside the interval $I$, but have exponentially decaying tails. In particular say 

\begin{align}
    &df_{j,j+1} = \frac{1}{2\sqrt{2\pi |I|^2}}e^{-\frac{j^2}{2|I|^2}} \\
    &dg_{j-1,j} = \frac{1}{\sqrt{2\pi |I|^2}}e^{-\frac{j^2}{2|I|^2}}
\end{align}
Therefore the magnitude of eq. \ref{eq: localized commutator} will go as $\exp(-2\pi^2 k^2 |I|^2/N^2)$. We note that this formula makes it seem that as one shrinks $|I|$ there will be no particle-hole creation, but this will not be the case as the formula does not hold for small $|I|$ as aliasing effects will dominate the discrete fourier transform. 

From eq. \ref{eq: localized commutator}, we can also see the effect of tuning the parameters to the free fermion radius ($U_0/J_0 = 1/16\pi^2$). Say we are at this radius and we consider the commutator of ${\mathcal{F}^I}^\dagger_{R}$ with $M_k$. Then around $k=0$ the commutator is low except very close to $k=0$ itself, but around $k=N$ it is large (except strictly near $k=N$). A similar argument can be made for the left handers. This is to say the following: if we create a single fermion of chirality $\nu$, it is at the free fermion radius that the localized operator ${\mathcal{F}^I}^\dagger_{\nu}$ will create the single fermion while also creating the fewest number of particle-hole pairs in the $\nu$ branch, but it comes at the cost of creating many particle-hole pairs in the branch opposite to $\nu$.

\end{document}